\documentclass[twocolumn,showpacs,preprintnumbers,amsmath,amssymb]{revtex4}

\usepackage{graphicx}
\usepackage{dcolumn}
\usepackage{bm}

\begin{document}

\preprint{}

\title{Spin Hall effect of light beam in left-handed materials}
\author{Hailu Luo}
\author{Shuangchun Wen}\email{scwen@hnu.cn}
\author{Weixing Shu}
\author{Zhixiang Tang}
\author{Yanhong Zou}
\author{Dianyuan Fan}
\affiliation{Key Laboratory for Micro/Nano Opto-Electronic Devices
of Ministry of Education, School of Computer and Communication,
Hunan University, Changsha 410082, China}
\date{\today}

\begin{abstract}
We establish a general propagation model to describe
the spin Hall effect of light beam in left-handed materials (LHMs).
A spin-dependent shift of the beam centroid perpendicular to the refractive index gradient
for the light beam through an air-LHM interface is demonstrated.
For a certain circularly polarized component, whether the transverse shift is positive or negative depends on
the magnitude of the refractive index gradient.
Very surprisingly, the spin Hall effect in the LHM is unreversed,
although the sign of refractive index gradient is reversed.
The physics underlying this counterintuitive effect
is that the spin angular momentum of photons is unreversed.
Further, we reveal that the angular shift in the LHM is reversed due to the negative diffraction.
These findings provide alternative evidence for that the linear momentum of photons is reversed,
while the spin angular momentum is unreversed in the LHM.
\end{abstract}


\maketitle

\section{Introduction}\label{SecI}
The spin Hall effect of light is the photonic version of the spin Hall effect
in electronic systems~\cite{Sinova2004,Murakami2003,Wunderlich2005},
in which the spin photons play the role of the spin charges, and a refractive index
gradient plays the role of the electric potential gradient~\cite{Onoda2004,Bliokh2006}.
Such a spatial gradient for the refractive
index could occur at an interface between two materials.
The spin Hall effect manifests itself as a spin-dependent shift of the beam centroid
perpendicular to the refractive index gradient
when photons passing through the interface.
For left and right circularly polarized beams, the
eigenvalues of the transverse shift are the same in magnitude but
opposite in direction. In fact, the translational inertial spin
effect has been predicted by Beauregard more than 40 years ago~\cite{Beauregard1965}.
This effect is significantly different from longitudinal Goos-H\"{a}nchen
shift~\cite{Goos1947} and transverse Imbert-Fedorov
shift~\cite{Fedorov1965,Imbert1972} in total internal reflection,
which are described as evanescent-wave penetration.
The splitting in the spin Hall effect, implied
by angular momentum conservation, takes
place as a result of an effective spin-orbit interaction.
In a glass-air interface, the spin-dependent transverse shifts are just a few tens of nanometers.
This is the reason why the tiny scale of the effect escaped detection for a long time.
More recently, the spin Hall effect has been detected in experiment
via quantum weak measurements~\cite{Hosten2008}.

The recent advent of negative-refraction materials~\cite{Shelby2001},
also known as left-handed materials (LHMs)~\cite{Veselago1968}, not only opens
a new way to generate the magnitude of refractive index gradient,
but also provides an unprecedented reverse of the sign of refractive index
at the interface of an ordinary right-handed material (RHM) and a LHM.
It is now conceivable that LHMs
can be constructed whose permittivity and
permeability values may be designed to vary
independently and arbitrarily, taking negative or near-zero value as
desired~\cite{Pendry2006}. Hence, it is possible for designing a
certain refractive index gradient to modulate the spin Hall effect.
The potential interests encourage us
to explore what would happen to the spin Hall effect of light beam in the air-LHM interface.
In the past several years many counterintuitive phenomena,
such as anomalous evanescent-wave amplification~\cite{Pendry2000},
unusual photon tunneling~\cite{Zhang2002}, negative Goos-H\"{a}nchen
shift~\cite{Berman2002},  reversed Doppler effect~\cite{Seddon2003}, and inverted Cherenkov
effect~\cite{Lu2003} in LHMs have been reported. Now a question naturally arise: Whether is the
the spin Hall effect reversed as expected? In addition, due to the different arguments on the direction of linear momentum in
LHMs~\cite{Riyopoulos2006,Kemp2007,Scalora2007,Yannopapas2008}, the issue of how
to describe photon momentum is still an open problem. Hence our another motivation is to
clarify whether the photon momentum is reversed in the LHM.
In principle, the reflection and transmission from the air-LHM interface
should fulfill the total momentum conservation law.
We believe that the study of the spin Hall effect may provide insights into the fundamental
properties of photon momentum in the LHMs.

In this work, we use an air-LHM interface as the refractive index
gradient to explore what would happen to the spin Hall effect of
light beam. First, starting from the representation of a plane-wave angular spectrum,
we establish a general beam propagation model to describe
the beam reflection and transmission.
Next, we uncover how the beam evolves, and how the spin Hall effect affects its
transverse shifts.  It is necessary to consider the spin Hall effect
in the presence of loss, since the loss is inherent to realizable LHMs.
For the purpose of comparison,
we also study the spin Hall effect in the air-RHM interface.
Then, we examine what roles the magnitude and the sign of the refractive
index gradient play in the spin Hall effect.
Finally, we explore whether the spin-dependent transverse shift
or diffraction-dependent angular shift is reversed. The transverse shifts and the angular shifts,
governed by the total momentum conservation law,
provide us a new way to clarify the nature of photon momentum in the LHMs.

\section{Beam reflection and transmission}\label{SecII}
\begin{figure}
\includegraphics[width=8cm]{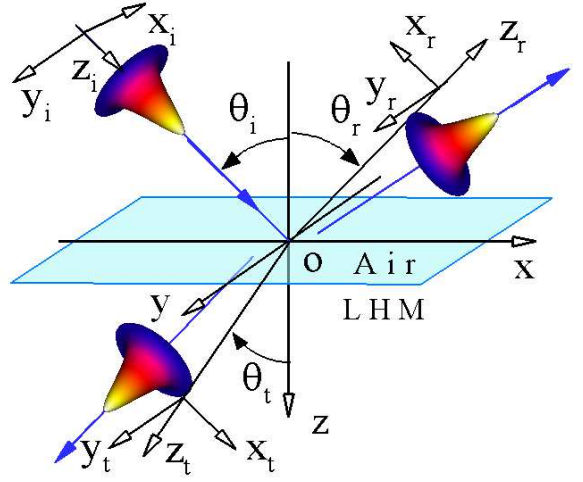}
\caption{\label{Fig1} (color online) Geometry of the beam reflection and
transmission from an air-LHM interface. The reflected and
transmitted beams will undergo the longitudinal shift, the
transverse shift, and the angular shift.}
\end{figure}

We begin to establish a general beam propagation model for describing
light beam reflection and refraction from a planar
air-LHM interface. Figure~\ref{Fig1} illustrates the Cartesian coordinate system. The
$z$ axis of the laboratory Cartesian frame ($x,y,z$) is normal to the air-LHM interface locate at $z=0$. We use the coordinate
frames ($x_a,y_a,z_a$) for individual beams, where $a=i,r,t$ denotes
incident, reflected, and transmitted beams, respectively. In the paraxial optics,
the incident field of an arbitrarily polarized beam can be written as
\begin{eqnarray}
\mathbf{E}_i(x_i,y_i,z_i
)&\propto&(\alpha\mathbf{e}_{ix}+\beta\mathbf{e}_{iy})
\exp\left[-\frac{k_0}{2}\frac{x_i^2+y_i^2}{z_{R}+
i z_i}\right].
\end{eqnarray}
Here $z_{R}= k_0 w_0^2 /2$ is the Rayleigh length
and $k_0=\omega/c$ is the wave number in the air. The coefficients
$\alpha$ and $\beta$ satisfy the relation
$\sigma=i(\alpha\beta^\ast-\alpha^\ast\beta)$. The polarization
operator $\sigma=\pm1$ corresponds to left- and right-handed
circularly polarized light, respectively. It is well known that circularly polarized
Gaussian beam can carry spin angular momentum $\sigma\hbar$ per
photon due to its polarization state~\cite{Beth1936}.

We first explore the reflected field, which can be solved by employing the Fourier
transformations. The complex amplitude can
be conveniently expressed as
\begin{eqnarray}
\mathbf{E}_r(x_r,y_r,z_r )&=&\int d k_{rx}dk_{ry}
\tilde{\mathbf{E}_r}(k_{rx},k_{ry})\nonumber\\
&&\times\exp [i(k_{rx}x_r+k_{ry}y_r+ k_{rz} z_r)],\label{nopr}
\end{eqnarray}
where $k_{rz}= \sqrt{k^2_0 - (k_{rx}^2+k_{ry}^2)}$. From a mathematical point of view, the approximate paraxial
expression for the field in Eq.~(\ref{nopr}) can be obtained by the expansion of the
square root of $k_{rz}$ to the first order~\cite{Lax1975}, which yields
\begin{eqnarray}
\mathbf{E}_r&=&\exp(i k_0 z) \int dk_{rx}dk_{ry}\tilde{\mathbf{E}_r}(k_{rx},k_{ry})\nonumber\\
&&\times\exp
\left[i\left(k_{rx}x_r+k_{ry}y_r-\frac{k_{rx}^2+k_{ry}^2}{2
k_0}z\right)\right].\label{par}
\end{eqnarray}
The reflected angular spectrum
$\tilde{E}_r(k_{rx},k_{ry})$ is related to the
boundary distribution of the electric field by means of the relation~\cite{Bliokh2006}
\begin{eqnarray}
\tilde{\mathbf{E}_r}=\left[
\begin{array}{cc}
r_p &\frac{k_{ry} \cot\theta_i}{k_0} (r_p+r_s) \\
-\frac{k_{ry} \cot\theta_i}{k_0} (r_p+r_s) & r_s
\end{array}
\right]\tilde{\mathbf{E}}_{p,s}^r,\label{matrixr}
\end{eqnarray}
where $r_p$ and $r_s$ are the Fresnel reflection coefficients. From
the Snell's law, we can get $k_{rx}=-k_{ix}$ and $k_{ry}= k_{iy}$.
Combining with boundary condition we get
\begin{equation}
\tilde{\mathbf{E}}_{p,s}^r\propto(\alpha\mathbf{e}_{rx}
+\beta\mathbf{e}_{ry})\exp\left[-\frac{z_R(k_{rx}^2+k_{ry}^2)}{2
k_0}\right].\label{asr}
\end{equation}
In fact, after the electric field on the plane $z_r=0$ is known,
Eq.~(\ref{par}) together with Eqs.~(\ref{matrixr}) and (\ref{asr}) provides the
expression of the field in the space $z_r>0$:
\begin{eqnarray}
\mathbf{E}_{r}&\propto&\bigg[\alpha r_p\left(1-i\frac{x_r}{z_R+ i
z_r}\frac{\partial \ln r_p}{\partial
\theta_i}\right)+i\beta\frac{y_r}{z_R+ i z_r}\nonumber\\
&&\times(r_p+r_s)\cot\theta_i
\bigg]\exp\left[-\frac{k_0}{2}\frac{x_r^2+y_r^2}{z_R+
i z_r}\right]\mathbf{e}_{rx}\nonumber \\
&&+\bigg[\beta r_s\left(1-i\frac{x_r}{z_R+ i z_r}\frac{\partial \ln
r_s}{\partial
\theta_i}\right)-i\alpha\frac{y_r}{z_R+ i z_r}\nonumber\\
&&\times(r_p+r_s)\cot\theta_i
\bigg]\exp\left[-\frac{k_0}{2}\frac{x_r^2+y_r^2}{z_R+ i
z_r}\right]\mathbf{e}_{ry}.\label{fieldr}
\end{eqnarray}
Note that the above expression of reflected field coincides
with the early results~\cite{Bliokh2007,Aiello2008,Aiello2009} for the reflection in air-RHM interface,
while the reflection coefficients are significantly different at the air-LHM interface.
Hence the spatial profile of the reflected beam is also evidently altered.

We next explore the transmitted field.
Similarly, the complex amplitude in the LHM can be
written as
\begin{eqnarray}
\mathbf{E}_t(x_t,y_t,z_t )&=&\int d k_{tx}dk_{ty}
\tilde{\mathbf{E}}_t(k_{tx},k_{ty})\nonumber\\
&&\times\exp [i(k_{tx}x_t+k_{ty}y_t)+i k_{tz} z_t],\label{nopt}
\end{eqnarray}
where $k_{tz}=-\sqrt{n^2 k_0^2 - (k_{tx}^2+k_{ty}^2)}$ and $n$ is the refractive index of the LHM.
The choice of negative sign of $k_{tz}$ ensures that power propagates away from the surface
to the $+z$ direction~\cite{Veselago1968}. In the paraxial approximation, we have
\begin{eqnarray}
\mathbf{E}_t&=&\exp(i n k_0 z) \int dk_{tx}dk_{ty}\tilde{\mathbf{E}_t}(k_{tx},k_{ty})\nonumber\\
&&\times\exp
\left[i\left(k_{tx}x_t+k_{ty}y_t-\frac{k_{tx}^2+k_{ty}^2}{2
n k_0}z\right)\right].\label{pat}
\end{eqnarray}
We note that the field of paraxial beams in LHMs can be written
in the similar way to that in RHMs,
while wave vector components undergo the negative refraction.

The transmitted angular spectrum
$\tilde{E}_t(k_{tx},k_{ty})$ in Eq.~(\ref{pat}) is related to the
boundary distribution of the electric field by means of the relation:
\begin{eqnarray}
\tilde{\mathbf{E}_t}=\left[
\begin{array}{cc}
t_p &\frac{k_{ty} \cot\theta_i}{k_0} (t_p-\eta t_s) \\
\frac{k_{ty} \cot\theta_i}{k_0} (\eta t_p-t_s)& t_s
\end{array}
\right]\tilde{\mathbf{E}}_{p,s}^t,\label{matrixt}
\end{eqnarray}
where $\eta=\cos\theta_t/\cos\theta_i$, $t_p$ and $t_s$ are the
Fresnel transmission coefficients. From the Snell's law under the
paraxial approximation, $k_{tx}=k_{ix}/\eta $ and $k_{ty}= k_{iy}$,
we obtain
\begin{equation}
\tilde{\mathbf{E}}_{p,s}^t\propto(\alpha\mathbf{e}_{tx}+\beta\mathbf{e}_{ty})
\exp\left[-\frac{z_{Rx}k_{tx}^2+z_{Ry} k_{ty}^2}{2n
k_0}\right].\label{ast}
\end{equation}
The interesting point we want to stress is that there are two
different Rayleigh lengths, $z_{Rx}=n\eta^2k_0w_0^2/2$ and
$z_{Ry}=n k_0w_0^2/2$, characterizing the spreading of the beam in
the direction of $x$ and $y$ axes, respectively~\cite{Luo2007}.
Substituting Eqs.~(\ref{matrixt}) and (\ref{ast}) into Eq.~(\ref{pat}),
we obtain the transmitted field in the space $z_t>0$:
\begin{eqnarray}
\mathbf{E}_{t}&\propto&\exp\left[-\frac{n
k_0}{2}\left(\frac{x_t^2}{z_{Rx}+ i z_t}+\frac{y_t^2}{z_{Ry}+ i
z_t}\right)\right]\nonumber \\&&\times\bigg[\alpha
t_p\left(1+i\frac{n \eta x_t}{z_{Rx}+ i z_t}\frac{\partial \ln
t_p}{\partial\theta_i}\right)\nonumber\\
&&+i\beta\frac{n y_t}{z_{Ry}+ i z_t}(t_p-\eta t_s)\cot\theta_i \bigg]\mathbf{e}_{tx}\nonumber\\
&&+\exp\left[-\frac{n k_0}{2}\left(\frac{x_t^2}{z_{Rx}+ i
z_t}+\frac{y_t^2}{z_{Ry}+ i
z_t}\right)\right]\nonumber\\
&&\times\bigg[\beta t_s\left(1+i\frac{n \eta x_t}{z_{Rx}+ i
z_t}\frac{\partial \ln
t_s}{\partial\theta_i}\right)\nonumber\\
&&+i\alpha \frac{n y_t}{z_{Ry}+ i z_t}(\eta t_p-t_s)\cot\theta_i
\bigg]\mathbf{e}_{ty}. \label{fieldt}
\end{eqnarray}
A further important point should be noted is that we have introduced
negative Rayleigh length or negative diffraction. The inherent physics underlying the
negative diffraction is the angular spectrum components undergo a negative phase
velocity~\cite{Luo2008a}.

\section{Spin Hall Effect}\label{SecIII}
In order to accurately describe the spin Hall effect, the issue of
loss of the LHM should be involved. Hence a certain dispersion relation, such as the
Lorentz medium model, should be introduced. The constitutive
parameters are
\begin{eqnarray}
\varepsilon(\omega)&=&1-\frac{\omega_{ep}^2}{\omega^2
-\omega_{eo}^2+i\omega \gamma_e},\\
\mu(\omega)&=&1-\frac{F\omega_{mp}^2}{\omega^2-\omega_{mo}^2+i\omega
\gamma_m}.
\end{eqnarray}
To avoid the trouble involving a certain value of frequency, we
assume the material parameters are
$\omega_{eo}=\omega_{mo}=\omega_o$, $\omega_{ep}=\omega_{mp}=\omega_o$, $F=1.52$,
and $\gamma_e=\gamma_m=0.001\omega_0$. Note that the Goos-H\"{a}nchen longitudinal
shift and Imbert-Fedorov transverse shift in such a
metamaterial have received much attentions~\cite{Berman2002,Menzel2008}.
Here we want to explore what would happen to the spin Hall effect of
light beam.

The intensity distribution of electromagnetic fields is
closely linked to the Poynting vector~\cite{Jackson1999} $I(x_a,y_a,z_a)\propto\mathbf{S}_a\cdot \mathbf{e}_{az}$.
Here the Poynting vector is given by $\mathbf{S}_a\propto\text{Re}[\mathbf{E}_a^\ast\times\mathbf{H}_a]$,
and the magnetic field can be obtained by $\mathbf{H}_a=-i \mu^{-1}\nabla\times\mathbf{E}_a$.
At any given plane $z_a=\text{const.}$, the beam centroid is given by~\cite{Aiello2008}
$\langle\mathbf{m}_a\rangle=\langle x_{a}\rangle \mathbf{e}_{ax} + \langle y_{a} \rangle \mathbf{e}_{ay}$, where
\begin{equation}
\langle\mathbf{m}_a\rangle = \frac{\int \int \mathbf{m}_a I(x_a,y_a,z_a)
\text{d}x_a \text{d}y_a}{\int \int I(x_a,y_a,z_a) \text{d}x_a \text{d}y_a}.\label{centroid}
\end{equation}
Because of the spin Hall effect, a linear polarized Gaussian beam
will be divided into two circularly polarized components with opposite shifts.

To illustrate the shifts, we now determine the centroid of the reflected beam.
Substituting Eq.~(\ref{fieldr}) into Eq.~(\ref{centroid}), we can obtain the
longitudinal and the transverse shifts. The longitudinal shift can be written as a sum of two terms
$D_x^r =\Delta x_r+\delta x_r$, and
\begin{eqnarray}
\Delta x_r &=&\frac{1}{k_0} \frac{\xi_p |r_p|^2 f_p^2
+\xi_s |r_s|^2  f_s^2 }{|r_p|^2  f_p^2 + |r_s|^2 f_s^2},\label{LSR}\\
\delta x_r &=& -\frac{z_r}{k_0 z_R} \frac{\rho_p |r_p|^2 f_p^2 +
\rho_s |r_s|^2 f_s^2 }{|r_p|^2 f_p^2 + |r_s|^2 f_s^2}.\label{LASR}
\end{eqnarray}
In a similar way, the transverse shift can also be divided into two terms
$D_y^r =\Delta y_r+\delta y_r$, we find
\begin{eqnarray}
\Delta y_r &=&-\frac{1}{k_0}\frac{f_p f_s \cot \theta_i }{|r_p|^2
f_p^2+|r_s|^2 f_s^2}\Bigl[\bigl(|r_p|^2 + |r_s|^2\bigr)\sin\psi\nonumber\\
&&+ 2 |r_p||r_s|\sin(\psi-\phi_p+\phi_s )\Bigr],\label{TSR}\\
\delta y_r &=&\frac{z_r}{k_0 z_R}\frac{f_p f_s  (|r_p|^2 -
|r_s|^2)\cot\theta_i \cos\psi}{|r_p|^2 f_p^2+|r_s|^2 f_s^2}.\label{TLSR}
\end{eqnarray}
Here $r_A=|r_A|\exp(i \phi_A)$, $A\in\{p, s\}$, $\alpha = f_p \in
\mathbf{R}$, $\beta = f_s \exp(i \psi)$,  $\rho_A = \mathrm{Re}
[{\partial \ln r_A}/{\partial \theta_i}]$, and $\xi_A=
\mathrm{Im}[{\partial \ln r_A}/{\partial \theta_i}]$.
From Eqs.~(\ref{LSR})-(\ref{TLSR}), we can find that both the longitudinal shift and the transverse shift can be written as
a combination of $z_r$-independent term and  $z_r$-dependent term.
Note that the $z_r$-independent term longitudinal shifts are reversed in the air-LHM interface,
which coincides well with the early prediction~\cite{Berman2002}.

The $z_r$-dependent term can be regarded as a small angler shift
inclining from the axis of beam centroid.
The corresponding longitudinal and transverse divergence angles are given by $\delta
x_r=z_r \Delta\theta_{rx}$ and $\delta y_r=z_r\Delta\theta_{ry}$, such that
\begin{eqnarray}
\Delta\theta_{rx}&=&-\frac{1}{k_0z_R} \frac{ \rho_p |r_p|^2
f_p^2 +\rho_s |r_s|^2 f_s^2 }{|r_p|^2 f_p^2 + |r_s|^2 f_s^2},\\
\Delta\theta_{ry} &=&\frac{1}{k_0 z_R}\frac{f_p f_s  (|r_p|^2 -
|r_s|^2)\cot\theta_i \cos\psi}{|r_p|^2 f_p^2+|r_s|^2 f_s^2}.
\end{eqnarray}
Here the longitudinal divergence angle is polarization independent,
while the transverse divergence angle depends on the polarization.
For a linear or a circularly beam,
the condition $\cos\psi=0$ is satisfied and the transverse divergence angle vanishes.

\begin{figure}
\includegraphics[width=8cm]{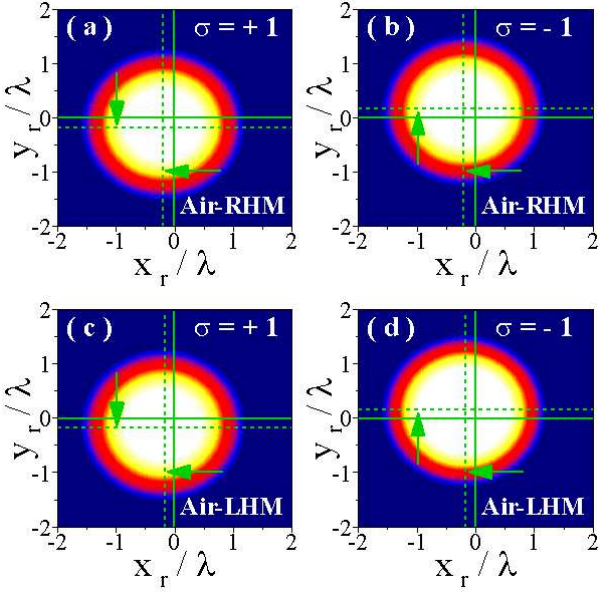}
\caption{\label{Fig2} (color online) The spin Hall effect induces
transverse shifts for the reflected beam at the interface
between air and low-refractive index medium.
(a), (b) Intensity distribution of the $\sigma=+1$ and $\sigma=-1$ beams reflected from the
air-RHM interface.
(c), (d) Intensity distribution of the $\sigma=+1$ and $\sigma=-1$ beams reflected from air-LHM interface.
The incident angle of the beam is choose as $\theta_i=\pi/6$.
The material parameters of LHM are $\omega=1.414\omega_0$, $\varepsilon=-1+0.003i$ and $\mu=-0.52+0.002i$.
To generate a reversed sign of refractive index gradient, the material parameters of
RHM are chosen as $\varepsilon=1+0.003i$ and $\mu=0.52+0.002i$.
The intensity distributions in the plane $z_r=z_R$ are plotted in normalized units.}
\end{figure}

To illustrate the spin Hall effect,
we consider a light beam incident from air to
a low-refractive-index medium.
Here the low-refractive-index medium means that
the values of the indices of the RHM and the LHM are less than the index of air.
In the case of reflection from the air-RHM interface, the left circularly polarized beam undergos
a negative transverse shift, while the
right circularly polarized beam exhibits a positive transverse shift [Figs.~\ref{Fig2}(a) and \ref{Fig2}(b)].
Compare to the transverse shifts in the air-RHM interface, we easily find that the corresponding shifts
in the air-LHM interface is unreversed [Figs.~\ref{Fig2}(c) and \ref{Fig2}(d)].
The unreversed transverse shifts mean that the spin Hall effect in the reflected beam is also unreversed.
In the absence of loss, the longitudinal shifts should vanish since the Fresnel coefficients are real~\cite{Bliokh2009}.
In our case, however, the longitudinal shifts present due to the loss inherent in the LHM.
Note that the longitudinal shifts in Fig.~\ref{Fig2} is unreversed in the LHM, because
the unreversed angular shift eliminate the reversed $z_r$-independent shift.

We now discuss the beam centroid of the transmitted beam.
After substituting Eq.~(\ref{fieldt}) into Eq.~(\ref{centroid}), we can obtain the
longitudinal shift and the transverse shift.
The longitudinal shift can be written as a sum of two terms
$D_x^t =\Delta x_t+\delta x_t$, then
\begin{eqnarray}
\Delta x_t &=& -\frac{\eta}{k_0} \frac{\zeta_p |t_p|^2 f_p^2
+\zeta_s |t_s|^2  f_s^2}{|t_p|^2 f_p^2 + |t_s|^2 f_s^2},\label{GHT}\\
\delta x_t &=&\frac{\eta z_t}{k_0 z_{Rx}} \frac{\varrho_p |t_p|^2 f_p^2
+\varrho_s |t_s|^2 f_s^2 }{|t_p|^2 f_p^2 + |t_s|^2 f_s^2}.\label{GHA}
\end{eqnarray}
In an analogous manner, the transverse shift can also be separated into two terms
$D_y^t =\Delta y_t+\delta y_t$, and
\begin{eqnarray}
\Delta y_t &=&-\frac{1}{k_0}\frac{f_p f_s \cot\theta_i }{|t_p|^2  f_p^2
+|t_s|^2 f_s^2}\bigl[\bigl(|t_p|^2 + |t_s|^2 \bigr)\sin\psi\nonumber\\
&&+2\eta|t_p||t_s|\sin(\psi - \varphi_p + \varphi_s ) \bigr],\\
\delta y_t &=&\frac{z_t}{k_0 z_{Ry}}\frac{f_p
f_s(|t_p|^2-|t_s|^2)\cot\theta_i\cos\psi}{|t_p|^2 f_p^2+|t_s|^2
f_s^2}.
\end{eqnarray}
Here $t_A=|t_A|\exp(i \varphi_A)$, $\varrho_A =\mathrm{Re}[
{\partial \ln t_A}/{\partial \theta_i}]$, and $\zeta_A=\mathrm{Im}[
{\partial \ln t_A}/{\partial \theta_i}]$.
It should be mentioned that $\delta x_t$ and $\delta y_t$ are given by functions of
Rayleigh lengths $z_{Rx}$ and $z_{Ry}$, respectively.
Hence, the angular shifts are purely diffraction effect.

For the transmitted beam, the longitudinal and the transverse divergence angles are given by $\delta
x_t=z_t \Delta\theta_{tx}$ and $\delta y_t=z_t \Delta\theta_{ty}$, we have
\begin{eqnarray}
\Delta\theta_{tx}&=&\frac{\eta}{k_0 z_{Rx}} \frac{\varrho_p |t_p|^2
f_p^2 +\varrho_s |t_s|^2 f_s^2 }{|t_p|^2 f_p^2 + |t_s|^2 f_s^2},\\
\Delta\theta_{ty} &=&\frac{1}{k_0 z_{Ry}}\frac{f_p
f_s(|t_p|^2-|t_s|^2)\cot\theta_i\cos\psi}{|t_p|^2 f_p^2+|t_s|^2
f_s^2}.
\end{eqnarray}
In principle, reflection and refraction of waves at the interface between two homogeneous
media is described by the Snell's law.
However, the angular divergences mean that the Snell's law cannot accurately describe such phenomena~\cite{Duval2006}.
Thus, the Snell's law need to be extended
to include spinning photons.

\begin{figure}
\includegraphics[width=8cm]{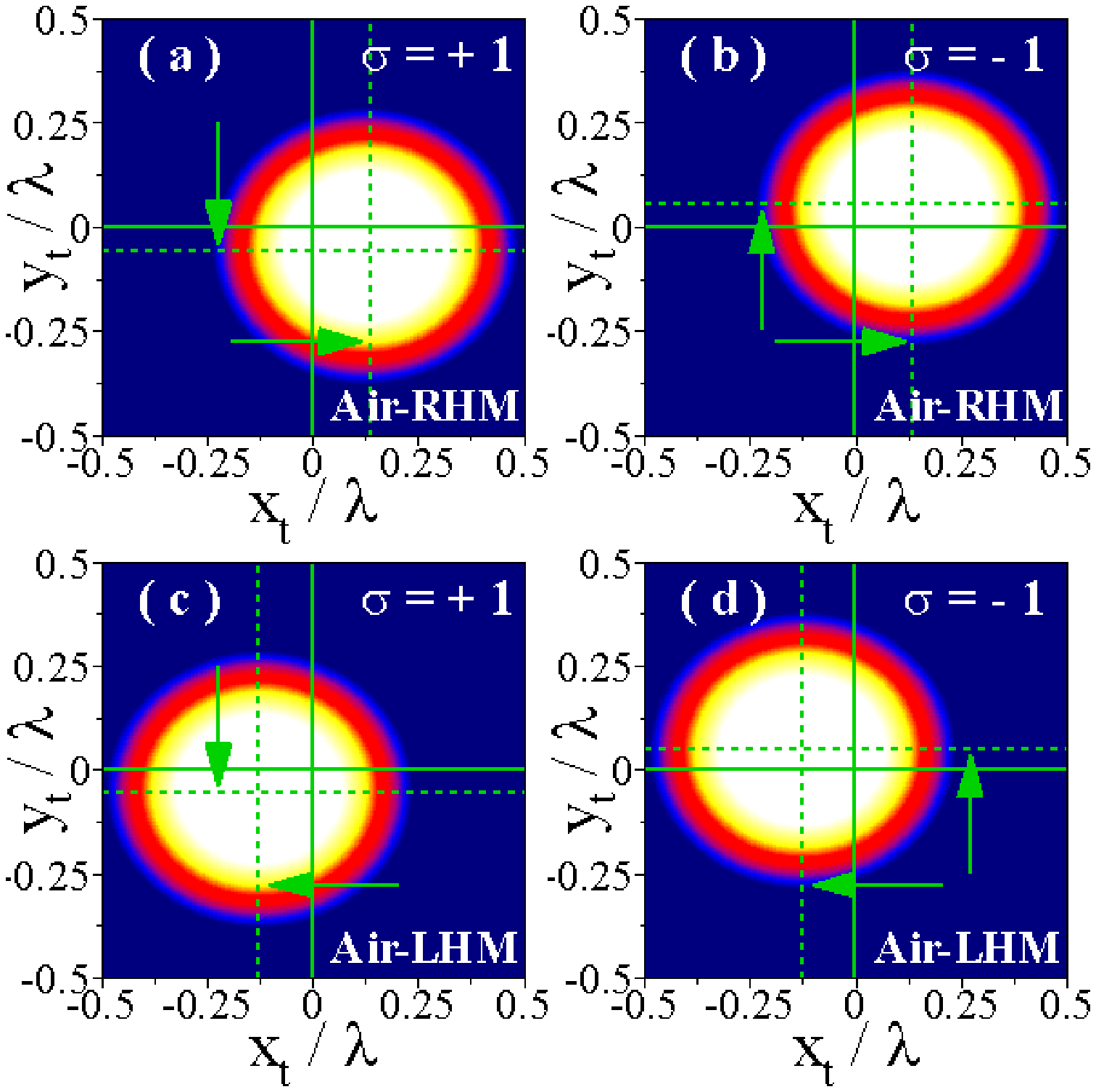}
\caption{\label{Fig3} (color online) The spin Hall effect induces
transverse shifts  for the transmitted beam from air to low-refractive-index medium.
(a), (b) Intensity distribution of the
transmission beam in the RHM for $\sigma=+1$ and $\sigma=-1$ components, respectively.
(c), (d) Intensity distribution of the transmitted beam in the LHM for $\sigma=+1$ and $\sigma=-1$ components, respectively.
The incident angle of the beam is choose as $\theta_i=\pi/6$. The material parameters are the same as in Fig.~\ref{Fig2}.
The intensity distributions in the plane $z_t=z_R$ are plotted in normalized units. }
\end{figure}

In the spin Hall effect of light, the refractive index gradient plays the role of the electric potential gradient.
Hence we attempt to examine what roles the magnitude and the sign of the refractive
index gradient play in the spin Hall effect.
We first consider the beam incident from air to a low-refractive-index medium.
For the left circularly polarized
component $\sigma=+1$, the transmitted field exhibits a negative transverse shift [Figs.~\ref{Fig3}(a) and \ref{Fig3}(c)].
For the right circularly polarized
component $\sigma=-1$, also presents a transverse shift, but in the opposite direction [Figs.~\ref{Fig3}(b) and \ref{Fig3}(d)].
Very surprisingly, the transverse shift caused by the spin Hall effect in the LHM is unreversed,
although the sign of refractive index gradient is reversed.
These results have also been proved by Krowne in his interesting work~\cite{Krowne2009}.
Furthermore we want to explore whether a reversed magnitude of refractive index gradient can
lead to a reversed spin Hall effect.

\begin{figure}
\includegraphics[width=8cm]{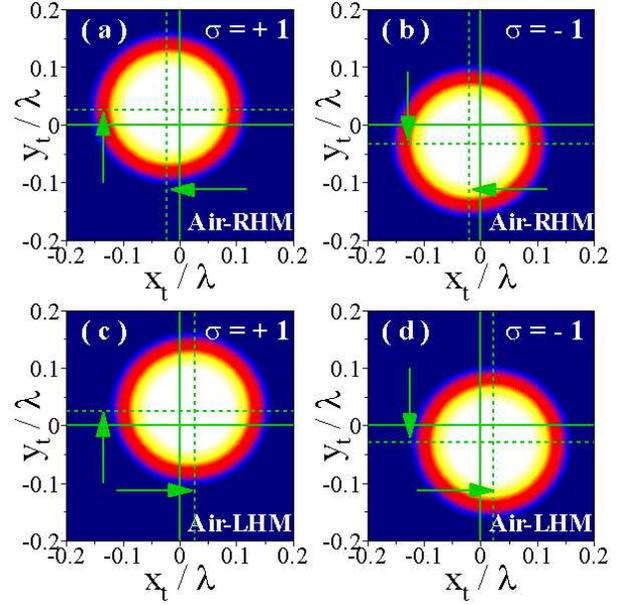}
\caption{\label{Fig4} (color online) The spin Hall effect induces
transverse shifts for the transmitted beam from air to high-refractive-index medium.
(a), (b) Intensity distribution of the
transmitted beam in the RHM for $\sigma=+1$ and $\sigma=-1$ components, respectively.
(c), (d) Intensity distribution of the transmitted beam in the LHM for $\sigma=+1$ and $\sigma=-1$ components, respectively.
The incident angle of the beam is choose as $\theta_i=\pi/6$.
To generate a high refractive index, the material parameters of LHM are $\varepsilon=-2+0.003i$ and $\mu=-1.52+0.002i$.
For comparison, the material parameters of
RHM are choose as $\varepsilon=2+0.003i$ and $\mu=1.52+0.002i$.
The intensity distributions in the plane $z_t= z_R$ are plotted in normalized units. }
\end{figure}

To generate a reversed magnitude of refractive index gradient,
we next investigate the beam incident from air to a high-refractive-index medium.
Here the high-refractive-index medium means
the values of the indices of the RHM and the LHM is larger than the index of air.
For the left circularly polarized
component $\sigma=+1$, the transmitted field exhibits a positive transverse shift [Figs.~\ref{Fig4}(a) and \ref{Fig4}(c)].
For the right circularly polarized
component  $\sigma=-1$, however, the transmitted field presents a negative transverse shift [Figs.~\ref{Fig4}(b) and \ref{Fig4}(d)].
We can find that the transverse shifts in the LHM are unreversed,
while the angular shifts is reversed because of the negative diffraction.
Comparing Fig.~\ref{Fig3} with Fig.~\ref{Fig4} shows that whether the spin Hall effect is reversed depends on
the magnitude the refractive index gradient but independs on its sign.

In order to reveal the physics underlying the unreversed spin Hall effect,
we focus on the momentum conservation laws which govern the beam reflection and transmission.
The monochromatic beam can be formulated as
a localized wave packet whose spectrum arbitrarily narrow~\cite{Bliokh2007}.
Let $a$th packet include $N_a$ photons, i.e., its field
energy is $W_a=N_a \omega$.
The linear momentum of the $a$th packet is
$\mathbf{p}_a=N_a\mathbf{k}_c^a$ (we set $\hbar=c=1$)
, and the conservation law for the linear momentum can be written as
$p_{x,y}^i=p_{x,y}^r+p_{x,y}^t$. For the $x$ component we have
$p_x^i=N_i k_i \sin\theta_{i}$,
$p_x^r=N_r k_r \sin\theta_r+N_r k_r \Delta\theta_{rx} \cos\theta_{r}$, and
$p_x^t=N_t k_t \sin\theta_t+N_t k_t \Delta\theta_{tx} \cos\theta_{t}$.
For the $y$ component we get
$p_y^i=0$,
$p_y^r=N_r k_r \Delta\theta_{rx}$, and
$p_y^t=N_t k_t \Delta\theta_{tx}$.
In the absence of loss, the total number of photons remain unchanged $N_r+N_t=N_i$.
The angular shifts should fulfill conservation law for $x$ and $y$ components of the
linear momentum:
\begin{eqnarray}
-Q_r \Delta\theta_{rx}\cos\theta_r+ n Q_t\Delta\theta_{tx}\cos\theta_t&=&0,\label{PX}\\
-Q_r \Delta\theta_{rx}+ n Q_t\Delta\theta_{ty}&=&0 \label{PY}.
\end{eqnarray}
Here $Q_r=N_r/N_i$ and $Q_t=N_t/N_i$ are the energy reflection and transmission coefficient, respectively.
In the frame of classic electrodynamics,
$Q_r=f_p^2 |r_p|^2+f_s^2|r_s|^2$ and $Q_t=n\eta(f_p^2 |t_p|^2+f_s^2|t_s|^2)/\mu$,
the conservation law, Eqs.~(\ref{PX}) and (\ref{PY}), still holds true.
These results coincide well with the linear momentum conservation laws in RHMs~\cite{Fedoseyev2009,Bliokh2009},
although the sign of refractive index is reversed.
From the linear momentum conservation law, we find that the reversed divergence angles
mean the linear momentum of photons should be reversed.

We proceed to consider the total angular momentum conservation law.
The $z$ component of total angular momentum of per one photon can be represented as a sum of the extrinsic orbital
angular momentum and intrinsic spin angular momentum~\cite{Bliokh2007}:
\begin{eqnarray}
j_{rz}&=&-\Delta y_r k_r \sin\theta_r+\sigma_r \cos\theta_r,\label{JRZ}\\
j_{tz}&=&-\Delta y_t k_t \sin\theta_t+\sigma_t \cos\theta_t.\label{JTZ}
\end{eqnarray}
The transverse shifts of the wave packet fulfill the conservation law for
the total angular momentum~\cite{Player1987,Fedoseyev1988}:
\begin{equation}
Q_r(S_{rz}-\Delta y_r k_r \sin\theta_r)+
Q_t(S_{tz}-\Delta y_t k_t \sin\theta_t)=S_{iz}.\label{totalam}
\end{equation}
The $z$ component of the spin angular momenta are given by
$S_{iz}=\sigma\cos\theta_i$, $S_{rz}=\sigma_r \cos\theta_r$,
and $S_{tz}=\sigma_t \cos\theta_t$ for incident beam, reflected beam, and transmitted beam, respectively.
The polarization degrees of the reflected and the transmitted beams are described by
\begin{eqnarray}
\sigma_r&=&\frac{2 f_p f_s|r_p||r_s|\sin[\psi - (\phi_p - \phi_s )]}{|r_p|^2  f_p^2
+|r_s|^2 f_s^2},\\
\sigma_t&=&\frac{2 f_p f_s|t_p||t_s|\sin[\psi - (\varphi_p - \varphi_s )]}{|t_p|^2  f_p^2
+|t_s|^2 f_s^2}.
\end{eqnarray}
In the regime of partial reflection and transmission, the Fresnel coefficients are real
($\phi_A=0$ and $\varphi_A=0$). Hence, the polarization degree
of the transmitted beam should be unreversed in the LHM.
The physics underlying the unreversed spin-dependent transverse shifts
is the unreversed spin angular momentum of photons.
A further point should mentioned that the unreversed angular momentum of photons
can provide a physical explanation of our early prediction:
The rotational Doppler effect in the LHM is unreversed,
although the linear Doppler effect is reversed~\cite{Luo2008b}.

To obtain a clear physical picture of the spin Hall effect,
we attempt to perform analyses on the $z$ component of the total angular momentum for each of
individual photons. The total angular momentum law for single photon is given by~\cite{Onoda2006}
\begin{equation}
-\Delta y_t k_t \sin\theta_t+\sigma_t \cos\theta_t=\sigma \cos\theta_i.
\end{equation}
When the photon incident from air to a low-refractive-index medium,
the incident angle is less than the refractive angle $\theta_i<|\theta_t|$.
The linear polarized beam can be represented as a superposition of equal $\sigma=+1$ and $\sigma=-1$ photons.
For the $\sigma=+1$ photons, the $z$ component of spin angular momentum $\sigma_t \cos\theta_t$ ($\sigma_t>0$) decreases
after entering the medium. Because of the conservation law, the total angular momentum must remain unchanged.
To conserve the total angular momentum, the photon must move to the direction $-y$ ($\Delta y_t<0$) [Fig.~\ref{Fig3}(a) and \ref{Fig3}(c)].
For the $\sigma=-1$ photons, the $z$ component of spin angular momentum $\sigma_t \cos\theta_t$ ($\sigma_t<0$) increases.
As the result, the photons must move to the direction $+y$ ($\Delta y_t>0$) [Fig.~\ref{Fig3}(b) and \ref{Fig3}(d)].
When the beam incident from air to a high-refractive-index medium,
the incident angle is less than the refractive angle $\theta_i>|\theta_t|$.
The spin Hall effect exhibit a reversed version as shown in Fig.~\ref{Fig4}.
This gives a simple way to understand how the light exhibits a spin Hall effect
and why the spin Hall effect is unreversed in the LHM.

For the spin Hall effect of light,
precise characterization requires measurement sensitivities at the angstrom
level~\cite{Hosten2008}. However, direct observation of the spin Hall effect is still remain an open challenge
in condensed matter physics~\cite{Sinova2004,Murakami2003,Wunderlich2005} and high-energy physics~\cite{Berard2006,Gosselin2007}.
Because of the close similarity in condensed matter,
high-energy physics, and optics,
the spin Hall effect of light will provide
indirect evidence in a diversity of physical systems~\cite{Bliokh2008}.
In general, introducing
the spin Hall effect into contemporary
photonics and nano-optics may result in the development of a
promising new area of research---spinoptics.
Recently, the technique of transformation optics has
emerged as a means of designing metamaterials that can
bring about unprecedented control of electromagnetic
fields~\cite{Pendry2006}. It is possible that the paths of different circularly polarized components
can be controlled by introducing a prescribed spatial variation
in the constitutive parameters.
Hence, the metamaterial is a good candidate to amplify the spin Hall effect.

\section{Conclusions}
In conclusion, we have established a general propagation model to describe
the spin Hall effect of light beam in the LHM.
The spin-dependent transverse shift of beam centroid perpendicular to the refractive index gradient
for the light beam through an air-LHM interface have been demonstrated.
For a certain circularly polarized component, whether the transverse shift is positive or negative depends on
the magnitude the refractive index gradient.
We have revealed that the spin Hall effect in the LHM is unreversed,
although the sign of refractive index gradient is reversed.
The physics underlying the unreversed spin Hall effect
is the unreversed spin angular momentum of photons.
We have found that the angular shifts in the LHM possesses a reversed version.
The inherent secret of this intriguing effect is the negative diffraction.
In the absence of loss, the angular shifts satisfy the total linear momentum conservation law,
and the transverse shifts fulfill the total angular momentum conservation law.
The study of the spin Hall effect will make a useful
contribution to clarify the nature of photons in the LHM.
Our results provide further evidence for that the linear momentum of photons is reversed,
while the spin angular momentum is unreversed in the LHM.

\begin{acknowledgements}
The authors are sincerely grateful to Professor A. B\'{e}rard for his helpful suggestions.
This research was supported by the National Natural Science
Foundation of China (Grants Nos. 10674045, 10804029, and 50802027).
\end{acknowledgements}

\end{document}